\documentclass[journal,twoside]{IEEEtran} 
\usepackage[dvips]{graphicx, color} 
\usepackage{amsmath}                
\usepackage{amssymb}
\usepackage{amsxtra}
\usepackage{bbm}
\usepackage{mathrsfs}
\usepackage{verbatim}             

\newtheorem{theorem}{Theorem}[section]
\newtheorem{lemma}[theorem]{Lemma}

\newcommand{\R} {\mathbb{R}}

\newcommand{\N} {\mathbb{N}}
\newcommand{\Z} {\mathbb{Z}}

\newcommand{\tP}{\mathrm{P}}
\newcommand{\tH}{\mathrm{H}}
\newcommand{\ul}[1]{\underline{#1} \!\,}
\newcommand{\cmpl} {\mathrm{c}}
\newcommand{\mat}[1]{\begin{pmatrix} #1 \end{pmatrix}}
\newcommand{\wti}[1] {\widetilde{#1}}
\newcommand{\genarg} {{\,\cdot\,}}

\newcommand{\msp}[1] {\mspace{#1 mu}}
\newcommand{\bels}[2] {
        \begin{equation} \label{#1} \begin{split} 
                #2 
        \end{split} \end{equation}
        }

\newcommand{\sett}[1] { \{ {#1} \} }
\newcommand{\setb}[1] { \bigl\{\, {#1} \,\bigr\} }
\newcommand{\abs}[1]{\lvert #1 \rvert}
\newcommand{\absb}[1]{\big\lvert #1 \big\rvert}
\newcommand{\pc}[2]{({#1} \,|\, {#2})}

\newcommand{\pb}[1]{\bigl({#1}\bigr)}
\newcommand{\pB}[1]{\Bigl({#1}\Bigr)}

\newcommand{\q}[1]{[{#1}]}

\newcommand{\qb}[1]{\bigl[{#1}\bigr]}
\newcommand{\qB}[1]{\Bigl[{#1}\Bigr]}
\newcommand{\qbb}[1]{\biggl[{#1}\biggr]}

\newcommand{\h}[1]{\{{#1}\}}

\newcommand{\hb}[1]{\bigl\{{#1}\bigr\}}

\newcommand{\pcb}[2]{\bigl({#1} \,\big|\, {#2}\bigr)}
\newcommand{\pcB}[2]{\Bigl({#1} \,\Big|\, {#2}\Bigr)}

\newcommand{\qc}[2]{[{#1} \,|\, {#2}]}
\newcommand{\qcb}[2]{\bigl[{#1} \,\big|\, {#2}\bigr]}
\newcommand{\qcB}[2]{\Bigl[{#1} \,\Big|\, {#2}\Bigr]}

\newcommand{\norm}[1]{\lVert #1 \rVert}
\newcommand{\normb}[1]{\big\lVert #1 \big\rVert}

\begin{document}

\title{Connection state overhead in a dynamic linear network}

\author{Oskari~Ajanki and~Antti~Knowles%
  \thanks{Oskari Ajanki is with the Department of
    Mathematics and Statistics, University of Helsinki 
    (email: \tt{oskari.ajanki@tkk.fi}).}
  \thanks{Antti Knowles is with the Institute for Theoretical Physics,
    ETH Z\"urich (email: \tt{aknowles@itp.phys.ethz.ch}).}}

\markboth{Submitted to IEEE Transactions on Information Theory}
{Ajanki and Knowles: Connection state overhead in a dynamic linear network}

\maketitle


\begin{abstract}
 
  We consider a dynamical linear network where nearest neighbours
  communicate via links whose states form binary (open/closed) valued
  independent and identically distributed Markov processes. 


  Our main result is the tight information-theoretic lower bound on
  the network traffic required by the connection state overhead, or
  the information required for all nodes to know their connected
  neighbourhood.

  These results, and especially their possible generalisations to more
  realistic network models, could give us valuable understanding of
  the unavoidable protocol overheads in rapidly changing Ad hoc and
  sensor networks.
\end{abstract}



\IEEEpeerreviewmaketitle


\section{Introduction}
\PARstart{I}{n} a dynamical network it is essential to keep track of
the connection state information in order to ensure efficient
transmission of data.  This requires additional information, in the
form a connection state overhead, to be sent through the network. For
networks with rapid dynamics (e.g.\ mobile networks) this overhead may
be large, and it is therefore of relevance to find some quantitative
measure of the required bandwidth.

In this paper we study a simple model of a one-dimensional network
introduced by Dey \cite{Prasenjit}, in which the links form identical,
independent and time-homogeneous discrete-time Markov processes in an
open/closed-binary space. In this case the required connectivity
information at a given node is simply the length of the path of open
links in either direction. The ensuing connection state overhead is
then quantified using information-theoretic methods. The relevant
quantity is the smallest possible number of bits per second required
for the connectivity overhead. Our main result is a sequence of upper
and lower bounds converging exponentially to this quantity, as well as
a simple and efficient method for their computation.

To our knowledge~\cite{Gallager} besides~\cite{Prasenjit} is the only
other work with the theme of quantifying the connection state overhead
by information theory.

The outline of the paper is as follows. In Section~\ref{section:
  Model} we introduce the network model and the connection state
variables. The overhead is quantified in Section~\ref{section: entropy
  rate}; we also introduce a sequence of bounds for this quantity,
derive an algorithm for their computation and show their exponential
convergence towards the exact optimal overhead cost.


\section{The model} \label{section: Model}

\begin{figure}[ht!]
\vspace{0.5cm}
\begin{center}
\includegraphics[width = 8cm]{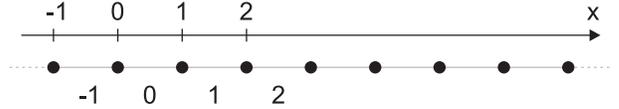}
\end{center}
\caption{\textit{The linear network. Nodes and links are indexed as shown.} 
\label{figure: network}}
\end{figure}

The one-dimensional network is composed of nodes and links connecting
neighbouring nodes. The nodes are labelled using the spatial variable
$x \in \Z$. We choose $x$ to increase to the right (see Figure
\ref{figure: network}). The links are labelled using the index $x \in
\Z$ such that the link $x$ connects the nodes $x$ and $x + 1$.  The
dynamics of the network is described using a discretised time variable
$ t \in \N $. The initial time is $t = 1$.

The probability space $\Omega := \{0,1\}^{\N \times \Z}$ contains
elements $\omega \in \Omega$ of the form $\omega = \{\omega_{tx} \,:\,
t \in \N,\,x \in \Z\}$. The state of a link $x$ at time $t$ is
described by the random variable $X_t(x)$ which is by definition equal
to $\omega_{tx}$; ``$1$'' stands for up or open, and ``$0$'' for down
or closed. We shall introduce a probability measure $\tP$ of $\Omega$
on the $\sigma$-field generated by the finite-dimensional cylindrical
subsets of $\Omega$.

All links $ x $ are assumed to have identical and independent
statistics: $\tP = \bigotimes_{x \in \Z} \mathrm{p}$ is a product over
each $x \in \Z$. We now consider $\mathrm{p}$, i.e.\ the time
evolution of a single link $x$. Since all links $x$ have identical
statistics, we consider only the the link $x = 1$ and write $X_t :=
X_t(1)$. The time evolution of $X := \{X_t \,:\, t \in \N\}$ (and
consequently of $X(x) := \{X_t(x) \,:\, t \in \N\}$) is given by an
autonomous~\footnote{We use the term 'autonomous' as a synonym for
  'time-homogeneous'.} Markov process. Using the abbreviation
\begin{equation*}
p(b\,|\,a) \;:=\; \tP [X_{t+1} = b \; | \; X_t = a]\,,
\end{equation*}
where $a,b \in \{0,1\}$, the distribution of the Markov process $ X $
is determined by the transition matrix
\begin{equation} \label{equation: markov matrix}
T \;=\; 
\left(\begin{IEEEeqnarraybox*}[][c]{,c/c,}
p(1\,|\,1)  &  p(1\,|\,0)  \\ 
p(0\,|\,1)  &  p(0\,|\,0)  
\end{IEEEeqnarraybox*}\right) 
\;:=\;
\left(\begin{IEEEeqnarraybox*}[][c]{,c/c,}
\overline{d} &  u  \\ 
          d  &  \overline{u} 
\end{IEEEeqnarraybox*}\right)
\,,
\end{equation}
where $u,d \in (0,1)$ are the free parameters of the model, and
$\overline{\lambda} := 1 - \lambda $ for any $\lambda \in [0,1]$. Thus
$ u $ (resp.\ $d$) is the probability that a closed (resp.\ open) link
is opened (resp.\ closed) after one time step.

The above Markov chain has a steady state probability distribution on
$\{0,1\}$. For $b \in \{0,1\}$ we have
\begin{equation*}
p(b) \;:=\; \lim_{t \to \infty} \tP[X_t = b \,|\, X_1 = a]\,,
\end{equation*}
regardless of the initial condition $a \in \{0,1\}$.
From above we get
\begin{subequations}
\begin{align}
U &:= p(1) = \frac{u}{u+d}\,, \\
D &:= p(0) = \frac{d}{u+d}\,.
\end{align}
\end{subequations}
Thus $U$ (resp.\ $D$) is the steady state probability of a link being
up (resp.\ down).

For simplicity we assume that at time $ 1 $ all the link variables
$\{X_1(x), \; x \in \Z\}$ are distributed according to the stationary
distribution. Autonomity of the links implies then that
\bels{stationary distribution}{
\tP[X_t(x) = 1] &\;:=\; U\,, \\
\tP[X_t(x) = 0] &\;:=\; D\,,
}
for all $ x \in \Z $ and $ t \in \N $. 
Note that this restriction can always be lifted since all our results
concern the limit $t \to \infty$. For any given initial distribution,
conditions \eqref{stationary distribution} will hold with arbitrary
accuracy for large enough times.

These remarks define $\tP$ uniquely. Figure \ref{figure: space-time
  diagram} shows a space-time diagram of a typical evolution of the
link variables.

\begin{figure}[ht!]
\vspace{0.5cm}
\begin{center}
\includegraphics[width = 8cm]{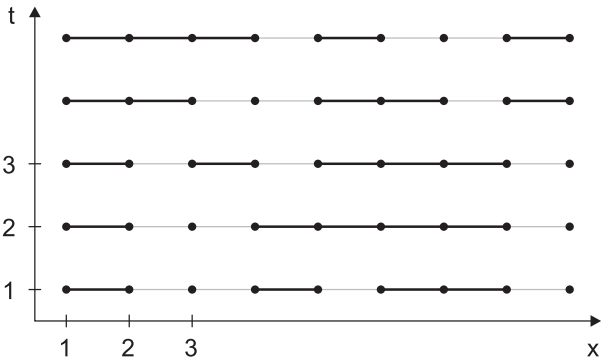}
\end{center}
\caption{\textit{A space-time view. Black links are open and gray links closed.} \label{figure: space-time diagram}}
\end{figure}


\subsection{Communications between the nodes}

We make the following assumptions about the communication capabilities
of the nodes.
\begin{itemize}
\item[(i)]
A node $ x $ is able to send a one message to its left neighbour $ x-1
$ and another (independent) message to its right member $ x+1 $ at
each time $ t $ via links $ x-1 $ and $ x $ respectively.
\item[(ii)]
If the link $ x $ is up at time $ t $, i.e., $ X_t(x) = 1 $, then the
nodes $ x $ and $ x+1 $ can receive the messages they have (possibly)
sent to each others at the previous time $ t-1 $. If the link $ x $ is
down at time $ t $ then these messages are lost. However, the nodes $
x $ and $ x+1 $ are able to observe that $ X_t(x) = 0 $ in this case.
\item[(iii)]
If a node $ x $ receives a message at time $ t $ it may resend it
immediately, i.e., the destination neighbour is able to receive the
message at the time $ t+1 $ provided the link between it and $ x $ is
up at $ t $.
\end{itemize}
Distant nodes are able to communicate by using the nodes between them
as relays. 
We assume that when a link is open it forms a communication
channel that has some finite transfer capacity. 
This last fact is not used for any calculations but is stated here to
make the subsequent considerations meaningful.


\subsection{The overhead messages}

In order to use efficient routing schemes it is important that a fresh
connectivity status of each node is known at all times.
Since the network is linear the relevant information is, for each node
$ x $, how far there exists an open path of links in both directions.
Because of the finite data propagation speed this connection state
information cannot be based on the current state of the network;
rather, it is extracted from the newest available data at $ x $ on
each link of the network. 
Since the network model is symmetric with respect to reflection about
$ x $ and the states of the links on left and right of $ x $ are
independent we may restrict ourselves to the right direction only. 
The quantity $ M_t(x) \in \N_0$~\footnote{We denote positive integers
  by $ \N $ and write $ \N_0 = \sett{0} \cup \N $ for non-negative
  integers.} expresses how many successive links are believed to be
open on the right-hand side of node $ x $ at time $t$. A natural
definition of $ M_t(x) $ in the light of the above remarks is then as
follows.

At the initial time $t = 1$ we set for all $ x \in \Z $
\begin{equation*}
M_1(x) := X_1(x)\,. 
\end{equation*}
As time advances nodes transmit information to their neighbours
according to the recursive scheme
\begin{equation} 
\label{recursion relation}
M_t(x) \,:=\, X_t(x) \, [M_{t - 1}(x+1) + 1]\,.
\end{equation}
Therefore
\begin{equation} 
\label{equation: expression for M}
M_t(x) \,=\, \sum_{m = 1}^t \prod_{k = 1}^m X_{t + 1 - k}(x - 1 + k)
\,,
\end{equation}
which, by the independence of the links, has a stationary distribution 
\begin{equation} \label{prob distribution of M}
\lim_{t \to \infty} \tP [M_t(x) = m] \;=\; D\, U^m
\,.
\end{equation}
Note that~\eqref{stationary distribution} implies that the equality
in~\eqref{prob distribution of M} holds even without the limit 
whenever $ t > m $.

\begin{figure}[ht!]
\vspace{0.5cm}
\begin{center}
\includegraphics[width = 6cm]{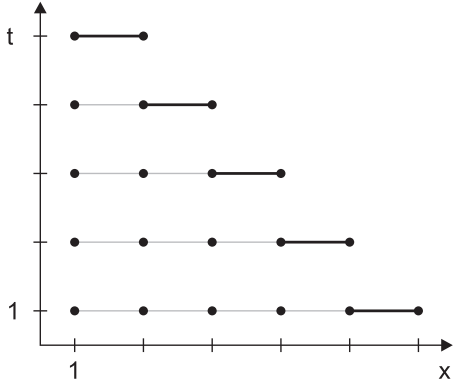}
\end{center}
\caption{\textit{The diagonal links $ \Lambda(t) $ contributing to
    $M_t$ (shown in black).}
\label{figure: space-time diagram 2}}
\end{figure}

Because of translation symmetry, we restrict ourselves to the studying
of the node $x = 1$ and abbreviate $M_t := M_t(1)$. Then
\eqref{equation: expression for M} becomes
\begin{equation} 
\label{equation: expression for M(1)}
M_t \,=\, \sum_{m = 1}^t \prod_{k = 1}^m X_{t + 1 - k}(k)\,.
\end{equation}
A glance at Figure \ref{figure: space-time diagram 2} shows that the
value of $M_t$ depends only on the link variables in the
time-space-diagonal $\Lambda(t) := \setb{ (s, x) \, : \; s =
  t + 1 - x \ge 1, x \ge 1} $.

To simplify notation we re-index link states on the diagonals 
$\Lambda(t) $, 
\begin{equation*}
Z_t(x) := 
\begin{cases}
X_{t + 1 - x}(x)\,, &x \leq t\,,
\\
0, & x > t
\,, 
\end{cases}
\end{equation*}
so that by~\eqref{equation: expression for M(1)} $ M_t $ becomes a
deterministic function of the infinite dimensional random vector
$\ul{Z}_t :=(Z_t(1),Z_t(2),\cdots) $.
Similarly, we define re-indexed messages on $ \Lambda(t) $ by setting
\begin{equation*}
\wti{M}_t(x) := M_{t + 1 - x}(x)
\,,
\end{equation*}
so that $ M_t(1) = M_1 = \wti{M}_t(1) $ and the recursion relation
\eqref{recursion relation} simplifies to 
\begin{equation} 
\label{recursion for Mtilde} 
\wti{M}_t(x) \,=\, Z_t(x) \, \bigl[\wti{M}_t(x+1) \,+\, 1 \bigr]
\,.
\end{equation}

Note that the effect of the transformation of the variables
$\sett{X_t(x)} \mapsto \sett{Z_t(x)} $ is equivalent to setting the
information propagation speed to infinity, as can be seen by comparing
the recursion relations \eqref{recursion relation} and
\eqref{recursion for Mtilde}. We may also consider a more general
network model in which each link $x$ transmits with a certain
(constant) speed $1/j(x)$, $j(x) = 0,1,2,3,\dots$. By a similar
variable transformation we can map this model to the infinite speed
model in the $Z_t(x)$ variables. Thus all following results are
equally valid for such more general networks.  
%
%
The relevance of the value $ M_t $ for the prediction of the true length of the open path for data sent at time $ t $ depends on the parameters $ u,v
$ (and of course $ j(x)$).


\subsection{Entropies related to the link variables}
The entropy of a single link (say $x = 1$, $t = 1$) is
\begin{equation*}
\tH (X_1) = h(U)\,,
\end{equation*}
where $\tH(\genarg)$ is the entropy functional on random variables and 
\begin{equation}
\label{auxiliary function h}
h(\lambda) := -\lambda \, \log \lambda - \overline{\lambda} \, \log \overline{\lambda}\,, \qquad \lambda \in [0,1]\,.
\end{equation}

The entropy rate of the process $X$ is by definition given by
\begin{equation*}
\mathscr{H}(X) := \lim_{t \to \infty} \frac{1}{t} \, \tH(X_t, \dots, X_1)\,.
\end{equation*}
Using the chain rule for entropy and Markovity (see \cite{Cover} for details) we may write
\begin{equation*}
\mathscr{H}(X) = \lim_{t \to \infty} \tH(X_{t + 1} \,|\, X_t)\,.
\end{equation*}
Since we assumed that $X_1$ is distributed according to the stationary distribution, we get
\begin{equation*}
\mathscr{H}(X) = \tH(X_2 \,|\, X_1)\,.
\end{equation*}
This may be easily evaluated to give
\begin{equation*}
\mathscr{H}(X) = U \, h(d) + D \, h(u)\,.
\end{equation*}

In the following we shall also encounter Markov chains $X^{(j)} =
\{X^{(j)}_t, \; t \in \N\}$ defined by
\begin{equation*}
X^{(j)}_t :=X_{jt}\,.
\end{equation*}
We therefore ``skip'' over $j$ links at each time step. The
corresponding transition probabilities are characterised by the two
off-diagonal elements of $T^j$, denoted by
\begin{subequations}
\begin{align}
u_j &\;:=\; \tP[X_{t + j} = 1 \; | \; X_t = 0]\,, \\
d_j &\;:=\; \tP[X_{t + j} = 0 \; | \; X_t = 1]\,. 
\end{align}
\end{subequations}

Precisely as above, we find for the entropy rate of this process:
\begin{equation} \label{j jump entropy for X}
\mathscr{H}\bigl(X^{(j)}) = U \, h(d_j) + D \, h(u_j)\,,
\end{equation}
where we used the fact that the stationary distribution of $X^{(j)}$
is the same as that of $ X $.


\section{Overhead cost: entropy rate of the overhead messages}
\label{section: entropy rate} We now quantify the optimal (i.e.\
smallest possible) cost of the connection state information overhead
by the entropy rate\footnote{Note that this must still be multiplied
  by two to account for both right and left directions.} of the
stochastic process $M := \{M_t,\; t \in \N\}$.  This corresponds to
the minimum amount of bits that need to be used on average to keep up
to date on the number of consecutive up-links in the right direction
from a fixed node $x$ (for more details see for instance \cite{Cover,
  Khinchin}). The rate is
\begin{align}
\mathscr{H}(M) &\,:=\, \lim_{t \to \infty} \frac{1}{t} \tH(M_t, \dots, M_1) 
\notag \\ 
\label{exact entropy rate}
&\;\,=\, \lim_{t \to \infty} \tH(M_t \,|\, M_{t - 1}, \dots, M_1)\,,
\end{align}
where the second equality follows by applying the chain rule of
entropy (note that both limits exist since $M$ is an autonomous
ergodic aperiodic process; see \cite{Cover} for details).


\subsection{Bounds for the message entropy rate}

The evaluation of \eqref{exact entropy rate} is tedious. A more
practical approach is to compute lower and upper bounds that can be
made as accurate as desired. Define for $j \in \N$
\begin{align} 
\label{equation: upper bound}
\mathscr{U}_j 
&\,:=\, \lim_{t \to \infty} \tH(M_t \,|\, M_{t - 1}, \dots, M_{t - j + 1})\,,
\\
\mathscr{L}_j 
&\,:=\, \lim_{t \to \infty} \tH(M_t \,|\, M_{t - 1}, \dots, M_{t - j + 1}, 
                              \ul{Z}_{t-j})
\,.
\end{align}
It should not come as a surprise that $\mathscr{U}_j$ (resp.\
$\mathscr{L}_j$) is an upper (resp.\ lower) bound for $\mathscr{H}(M)$
that becomes arbitrarily accurate in the limit $j \to \infty$. This is
the content of the following.

\begin{lemma} 
\label{lem: bounds}
The sequence $\{\mathscr{U}_j\}_{j \in \N}$ is non-increasing and
$\{\mathscr{L}_j\}_{j \in \N}$ is nondecreasing. Furthermore for all
$j \in \N$ we have
\begin{equation*}
\mathscr{L}_j \,\leq\, \mathscr{H}(M) \,\leq\, \mathscr{U}_j\,.
\end{equation*}
Finally,
\begin{equation*}
\mathscr{U}_j - \mathscr{L}_j \;\leq\; C \, \abs{1-u-d}^j
\,,
\end{equation*}
for some constant $ C = C(u,v) $. 
\end{lemma}

\begin{proof}
  We omit the (easy) proof of monotonicity of the sequences as well as
  the fact that they are bounds for $\mathscr{H}(M)$ (see for instance
  Lemma 4.4.1 in~\cite{Cover}). The convergence of the bounds is postponed to
  Theorem~\ref{thm: convergence}, as it is easiest to prove using
  results from the following section.
\end{proof}

\subsection{A recursive scheme for the bounds} 
\label{sect: recursion}

In this section we derive the main result: A recursive algorithm for
computing the bounds $\mathscr{L}_j$, $\mathscr{U}_j$ and thus for
approximating the exact entropy rate $\mathscr{H}(M)$ to an arbitrary
accuracy.

For the proof it will be useful to rewrite the entropy by partitioning
the probability space $\Omega$. Let $A \subset \Omega$ be an event.
Define $\tH(\genarg \,:\, A)$ as the entropy functional computed using
the conditional probability measure $ \tP[\genarg \,|\, A]
$. 
For two random variables $X,Y$ we have, for example,
\begin{align*}
&\tH (X \,|\, Y \,:\, A) \\
&=\; - \sum_{x,y} \; \tP[X=x , Y=y \,|\, A] \, \log \tP[X=x \,|\, Y=y , A]\,.
\end{align*}

\begin{lemma} \label{lemma: partitioning the entropy} If $ A $ lies in
  the $\sigma$-field $ \sigma(X,Y) $ generated by $(X,Y)$, then
\begin{align}
\tH(X \,|\, Y) \;=\; \tH(I_A \,|\, Y) \,+\, \notag
&\tP[A] \, \tH(X \,|\, Y \,:\, A) \,\\ +\, 
&\tP[A^\cmpl] \, \tH(X \,|\, Y \,:\, A^\cmpl)
\label{equation: partitioning the entropy} 
\,,
\end{align}
where $I_A$ is the indicator function of the event $ A $, and $
A^\cmpl $ denotes the complement of the set $ A $.
\end{lemma}
Note that if $A \in \sigma(Y)$ the first term of \eqref{equation:
  partitioning the entropy} vanishes.
\begin{proof}
  Using the fact that $ I_A $ is a deterministic function of $ (X,Y) $ as
  well as the chain rule we have
\begin{align*}
\tH(X \,|\, Y) \;&=\; \tH(X,Y, I_A \,|\, Y)
\;=\; \tH(X, I_A \,|\, Y) \\
&=\; \tH(I_A \,|\, Y) + H(X \,|\, I_A, Y)\,.
\end{align*}
The second term is equal to
\begin{align*}
&-\! \sum_{i \in \{0,1\}}\sum_{x,y}\, \tP[X = x, Y = y, I_A = i] \,\notag\\ 
&\msp{100} \log \tP[X = x \,|\, Y=y, I_A = i] \notag \\
&=\; - \sum_{i \in \{0,1\}} \tP[I_A = i] \sum_{x,y} \; \tP[X = x, Y = y \,|\, I_A = i] \, \notag\\
&\msp{177} \log \tP[X = x \,|\, Y=y, I_A = i]
\notag \\
&=\, \tP[A] \, \tH(X \,|\, Y \,:\, A) + \tP[A^\cmpl] \, \tH(X \,|\, Y \,:\, A^\cmpl)
\,.
\end{align*}
\end{proof}

We now introduce two sequences that will play a key role in the
following. For $ j \in \N$ define
\begin{equation*}
p_j \;:=\; \lim_{t\to \infty}\, \tP \bigl[M_t > \max \sett{ M_{t-1}, \dots, M_{t-j}}\bigr]
\,;
\end{equation*}
we also set $p_0 := 1$. Define furthermore the differences
\begin{equation*}
r_j \;:=\; p_{j - 1} - p_j\,,
\end{equation*}
for $j \in \N$.

In order to avoid writing explicit limits in the following we
introduce the equivalence relation $\sim$ to denote asymptotic
equality: $a(t) \sim b(t)$ means $\lim_{t \to \infty} a(t) = \lim_{t
  \to \infty} b(t)$.

\begin{theorem} 
\label{thm: main result}
The sequence of bounds $ \mathscr{L}_j $, $ \mathscr{U}_j $ can be
computed recursively from
\begin{subequations}
\label{the recursion relation}
\begin{align}  
\label{the recursion relation of lower bounds}
\mathscr{L}_{j+1} &\;=\; \mathscr{L}_j \,+\;\frac{p_j}{D} \, \Bigl[ \mathscr{H}\bigl(X^{(j+1)}\bigr) - \mathscr{H}\bigl(X^{(j)}\bigr) \Bigr]\,, 
\\ 
\label{the recursion relation of upper bounds}
\mathscr{U}_{j+1} &\;=\; \mathscr{L}_j \,+\;\frac{p_j}{D} \, \Bigl[ \tH (X_1) - \mathscr{H}\bigl(X^{(j)}\bigr)\Bigr]
\,, 
\end{align}
\end{subequations}
and
\begin{subequations} 
\label{initial value of recursion}
\begin{align}
\label{initial value of recursion a}
\mathscr{L}_1 \;=&\; \frac{1}{D} \mathscr{H}\bigl(X^{(1)}\bigr)\,, \\
\mathscr{U}_1 \;=&\; \frac{1}{D} \tH (X_1)\,.
\end{align}
\end{subequations}
\end{theorem}

Note that the probabilities $p_j$ (or, equivalently, the differences
$r_j$) must still be computed; this is done in Appendix \ref{sect:
  calculation of p_j}.  Everything else in the above expressions is
known: $\mathscr{H}\bigl(X^{(j+1)}\bigr)$ was computed in \eqref{j
  jump entropy for X}, and $\tH (X_1) = h(U)$.

A direct consequence of the theorem is an expression for the exact
entropy rate: From \eqref{initial value of recursion a} and \eqref{the
  recursion relation of lower bounds} we get
\begin{equation} 
\label{exact entropy rate as a sum over j}
\mathscr{H}(M) \;=\; \frac{1}{D} \, \sum_{j = 1}^\infty \; r_j \, \mathscr{H}\bigl(X^{(j)}\bigr)\,.
\end{equation}

\begin{proof}[Theorem \ref{thm: main result}]
We first introduce some notation. Define the vector
\begin{equation}
\ul{M}_{t - 1}^{(j)} \,:=\, (M_{t - 1}, \dots, M_{t - j})
\end{equation}
and the $\infty$-norm $\abs{\genarg}$ defined by 
\[
\abs{(m_1, \dots, m_j)} \,:=\, \max \sett{ m_1, \dots, m_j }
\,.
\]

The key idea of the proof is to partition the probability space
$\Omega = A \cup A^\cmpl$, where
\begin{equation*}
A \,:=\, \setb{M_t > \absb{\ul{M}_{t - 1}^{(j)}}}\,,
\end{equation*}
and use Lemma \ref{lemma: partitioning the entropy}. Some of the rigorous
proofs of the intuitively plausible steps (a-f) are postponed to Lemma
\ref{lemma: proof of steps}. We have
\begin{align*}
\mathscr{L}_{j+1} \;\sim\;&\tH \pcb{M_t}{\ul{M}_{t - 1}^{(j)}, \ul{Z}_{t-j-1}}
\notag \\
\overset{\text{(a)}}{=}\;&\tH \pcb{I_A}{\ul{M}_{t - 1}^{(j)}, \ul{Z}_{t-j-1}} 
\notag \\
&+\,\tP[A] \, \tH \pcb{M_t}{\ul{M}_{t - 1}^{(j)}, \ul{Z}_{t-j-1} \,:\, A}
\notag \\
&+\,\tP[A^\cmpl] \, \tH \pcb{M_t}{\ul{M}_{t - 1}^{(j)}, \ul{Z}_{t-j-1} \,:\, A^\cmpl}
\notag \\
\overset{\text{(b)}}{=}\;&\tH \pcb{I_A}{\ul{M}_{t - 1}^{(j)}, \ul{Z}_{t-j-1}}
\notag \\
&+\,\tP[A] \, \tH \pcb{M_t}{\ul{M}_{t - 1}^{(j)}, \ul{Z}_{t-j-1} \,:\, A}
\notag \\
&+\,\tP[A^\cmpl] \, \tH \pcb{M_t}{\ul{M}_{t - 1}^{(j-1)}, \ul{Z}_{t-j} \,:\, A^\cmpl}
\notag \\
\overset{\text{(c)}}{=}\;&\tH \pcb{M_t}{\ul{M}_{t - 1}^{(j-1)}, \ul{Z}_{t-j}}
\notag \\
&+\,\tH \pcb{I_A}{\ul{M}_{t - 1}^{(j)}, \ul{Z}_{t-j-1}}
\,-\,\tH \pcb{I_A}{\ul{M}_{t - 1}^{(j-1)}, \ul{Z}_{t-j}}
\notag \\
&+\, \tP[A] \, 
\begin{aligned}[t]
\Bigl[\,&\tH \pcb{M_t}{\ul{M}_{t - 1}^{(j)}, \ul{Z}_{t-j-1} \,:\, A} \\
&-\, \tH \pcb{M_t}{\ul{M}_{t - 1}^{(j-1)}, \ul{Z}_{t-j} \,:\, A}\,\Bigr]
\end{aligned}
\notag \\
\overset{\text{(d)}}{\sim}\;&\mathscr{L}_j + 0 + \tP [A] \,
\qB{\tH \pcb{M_t}{\ul{Z}_{t-j-1}} - 
\tH \pcb{M_t}{\ul{Z}_{t-j}}}
\notag \\
\overset{\text{(f)}}{\sim}\;&\mathscr{L}_j + \frac{p_j}{D} \,
\Bigl[\mathscr{H}\pb{X^{(j+1)}} - 
\mathscr{H}\bigl(X^{(j)}\bigr)\Bigr]\,,
\notag
\end{align*}
where (a) follows from Lemma \ref{lemma: partitioning the entropy};
(b) from Lemma \ref{lemma: proof of steps} (i); (c) from Lemma
\ref{lemma: partitioning the entropy} applied to $ X = M_t $ and $ Y =
(\ul{M}_{t - 1}^{(j-1)}, \ul{Z}_{t-j}) $; (d) from Lemma \ref{lemma:
  proof of steps} (ii),(iv),(v); and (f) from Lemma \ref{lem: first
  order bounds}.

Similarly,
\begin{align*}
\mathscr{U}_{j+1} \;\sim\;&\tH \pcb{M_t}{\ul{M}_{t - 1}^{(j)}}
\notag \\
\overset{\text{(a)}}{=}\;&\tH \pcb{I_A}{\ul{M}_{t - 1}^{(j)}}
\begin{aligned}[t]
\,+\,&\tP[A] \, \tH \pcb{M_t}{\ul{M}_{t - 1}^{(j)} \,:\, A}
\\
\,+\,&\tP[A^\cmpl] \, \tH \pcb{M_t}{\ul{M}_{t - 1}^{(j)} \,:\, A^\cmpl}
\end{aligned}
\notag \\
\overset{\text{(b)}}{=}\;&\tH \pcb{I_A}{\ul{M}_{t - 1}^{(j)}}
\begin{aligned}[t]
\,+\, &\tP[A] \, \tH \pcb{M_t}{\ul{M}_{t - 1}^{(j)} \,:\, A}
\\
\,+\, &\tP[A^\cmpl] \, \tH \pcb{M_t}{\ul{M}_{t - 1}^{(j-1)}, \ul{Z}_{t-j} \,:\, A^\cmpl}
\end{aligned}
\notag \\
\overset{\text{(c)}}{=}\;&\tH \pcb{M_t}{\ul{M}_{t - 1}^{(j-1)}, \ul{Z}_{t-j}}
\notag \\
&+\,\tH \pcb{I_A}{\ul{M}_{t - 1}^{(j)}}
\,-\,\tH \pcb{I_A}{\ul{M}_{t - 1}^{(j-1)}, \ul{Z}_{t-j}}
\notag \\
&+\,\tP[A] \, 
\begin{aligned}[t]
\Bigl[\,&\tH \pcb{M_t}{\ul{M}_{t - 1}^{(j)} \,:\, A} \\
& -\,\tH \pcb{M_t}{\ul{M}_{t - 1}^{(j-1)}, \ul{Z}_{t-j} \,:\, A}\,\Bigr]
\end{aligned}
\\ \overset{\text{(d)}}{\sim}\; 
&\mathscr{L}_j + 0 + \tP [A] \,
\qB{\tH \pb{M_t} - 
\tH \pcb{M_t}{\ul{Z}_{t-j}}}
\\ \overset{\text{(f)}}{\sim}\; 
&\mathscr{L}_j + \frac{p_j}{D} \,
\qB{\tH(X_1) - 
\mathscr{H}\pb{X^{(j)}}}
\,,
\end{align*}
where (a) follows from Lemma \ref{lemma: partitioning the entropy};
(b) from Lemma \ref{lemma: proof of steps} (i); (c) from Lemma
\ref{lemma: partitioning the entropy}; (d) from Lemma \ref{lemma:
  proof of steps} (ii),(iii),(v); and (f) from Lemma \ref{lem: first
  order bounds}.

The initial values \eqref{initial value of recursion} follow from
Lemma \ref{lem: first order bounds}.
\end{proof}

\begin{lemma} 
\label{lemma: proof of steps}
Using the notation of the proof of Theorem \ref{thm: main result}, we have
\begin{align*}
\text{(i)} \;\;\;
&\tH \pcb{M_t}{\ul{M}_{t - 1}^{(j)} \,:\, A^\cmpl} 
\begin{aligned}[t]
&= \, \tH \pcb{M_t}{\ul{M}_{t - 1}^{(j)}, \ul{Z}_{t-j-1} \,:\, A^\cmpl}
\\
&= \, \tH \pcb{M_t}{\ul{M}_{t - 1}^{(j-1)}, \ul{Z}_{t-j} \,:\, A^\cmpl}\,,
\end{aligned}
\\
\text{(ii)} \;\;\;
&\tH \pcb{I_A}{\ul{M}_{t - 1}^{(j)}}
\begin{aligned}[t]
&= \,\tH \pcb{I_A}{\ul{M}_{t - 1}^{(j)}, \ul{Z}_{t-j-1}} 
\\
&=\, \tH \pcb{I_A}{\ul{M}_{t - 1}^{(j-1)}, \ul{Z}_{t-j}}\,,
\end{aligned}
\\
\text{(iii)} \;\;\;
&\tH \pcb{M_t}{\ul{M}_{t-1}^{(j)} \,:\, A} \,\sim\, \tH(M_t)\,,
\\
\text{(iv)} \;\;\;
&\tH \pcb{M_t}{\ul{M}_{t - 1}^{(j)}, \ul{Z}_{t-j-1} \,:\, A}
\,\sim\, \tH \pcb{M_t}{\ul{Z}_{t-j-1}} \,,
\\
\text{(v)} \;\;\;
&\tH \pcb{M_t}{\ul{M}_{t - 1}^{(j-1)}, \ul{Z}_{t-j} \,:\, A}
\,\sim\, \tH \pcb{M_t}{\ul{Z}_{t-j}}
\,.
\end{align*}
\end{lemma}
The proof of Lemma \ref{lemma: proof of steps} is banished to Appendix
\ref{app: proof of steps}. To complete the proof of Theorem \ref{thm:
  main result} we still need the first order bounds $\mathscr{L}_1$,
$\mathscr{U}_1$.

\begin{lemma} 
\label{lem: first order bounds}
For any $j \in \N$ we have
\begin{equation} 
\label{first order lower bound}
\lim_{t \to \infty} \tH(M_t \,|\, \ul{Z}_{t-j}) 
\,=\, \frac{1}{D} \mathscr{H} \pb{X^{(j)}}\,.
\end{equation}
In particular, $\mathscr{L}_1 = \mathscr{H}(X)/D$. 
Furthermore,
\begin{equation*}
\lim_{t \to \infty} \tH(M_t) = \frac{1}{D} \, \tH(X_1)\,,
\end{equation*}
so that $\mathscr{U}_1 = \tH (X_1) / D$.
\end{lemma}

\begin{proof}
  By the recursion relation \eqref{recursion for Mtilde} we have $M_t
  = Z_t(1) \cdot \qb{\wti{M}_t(2) + 1}$.  Since $\wti{M}_t(2) \geq 0$
  we have by bijectivity and the chain rule
\begin{align*}
&\tH (M_t \,|\, \ul{Z}_{t-j}) \\
&=\;\tH \pcb{Z_t(1), Z_t(1) \cdot \q{\,\wti{M}_t(2) + 1\,}}{\ul{Z}_{t-j}}
\\
&=\;\tH\pcb{Z_t(1)}{\ul{Z}_{t-j}} + \tH \pcb{Z_t(1) \cdot \q{\,\wti{M}_t(2) + 1\,}}{Z_t(1),\ul{Z}_{t-j}}\\
&=\;\mathscr{H} \pb{X^{(j)}} \,+\,\tP \qb{Z_t(1) = 1} 
\\
&\msp{40}\cdot\, \tH \pcB{Z_t(1) \cdot \q{\,\wti{M}_t(2) + 1\,}}{\ul{Z}_{t-j} \,:\, \h{Z_t(1) = 1}} 
\\
&\msp{25}+\,\tP \qb{Z_t(1) = 0} \\ 
&\msp{40}\cdot\, \tH \pcB{Z_t(1) \cdot \q{\,\wti{M}_t(2) + 1\,}}{\ul{Z}_{t-j} \,:\, \h{Z_t(1) = 0}}
\\
&=\; \mathscr{H} \pb{X^{(j)}} + U \, \tH \pcb{\wti{M}_t(2)}{\ul{Z}_{t-j} \,:\, \h{Z_t(1) = 1}}
\\
&=\; \mathscr{H} \pb{X^{(j)}} + U \, \tH \pcb{\wti{M}_t(2)}{\ul{Z}_{t-j}}\,,
\end{align*}
where the last step follows from the fact $\wti{M}_t(2)$ is
independent of $Z_t(1)$. Using translation invariance we therefore get
\begin{equation*}
\lim_{t \to \infty} \tH (M_t \,|\, \ul{Z}_{t-j}) \,=\, \mathscr{H} \pb{X^{(j)}}
 + U \lim_{t \to \infty} \tH (M_t \,|\, \ul{Z}_{t-j})\,,
\end{equation*}
and \eqref{first order lower bound} follows.

Furthermore from $ \tP [M_t = m] \sim U^m \, D $ we get
\begin{align*}
\lim_{t \to \infty} \tH (M_t) \;
&=\, - \sum_{m = 0}^\infty \; U^m \, D \, \log \pb{U^m \, D } \\
&=\; \frac{h(U)}{D} \;=\; \frac{\tH(X_1)}{D}\,,
\end{align*}
where the function $ h $ is defined in~\eqref{auxiliary function h}. 
\end{proof}


\subsection{Convergence of the bounds}
We now address the convergence of the bounds, thus completing the
proof of Lemma \ref{lem: bounds}.
\begin{theorem}
\label{thm: convergence}
  For any $ u,v \in (0,1) $ there exists a constant $C = C(u,d) < \infty $
  such that
\begin{equation*}
\mathscr{U}_j - \mathscr{L}_j \;\leq\; C \, \abs{1 - u - d}^j \,.
\end{equation*}
\end{theorem}
\begin{proof}
  We start with three auxiliary results.

  First, we notice that the eigenvalues of the single link transition
  matrix $ T $ in~\eqref{equation: markov matrix} are $ 1, 1 - u - d
  $, and since $ \abs{1-u-d} < 1 $ the limit
\begin{equation}
T_* \;:=\; \lim_{j \to \infty}T^{j} \;=\; 
\mat{U & U \\D & D}
\,,
\end{equation}
exists. (This is just a restatement that the $ X $ has a unique stationary
distribution.) The convergence is exponentially fast, i.e.,
\begin{equation}\label{matrix convergence}
\normb{ T^j - T_*} \;\leq\; k_1\, \abs{1 - u - d}^j 
\,,
\end{equation}
where $\norm{\genarg}$ is a matrix norm and $ k_1 = k_1(u,v) $ is some
finite constant.

Second, the smooth function $ g $ on $ 2 \times 2 $ matrices $
(0,1)^{2\times 2}$, defined by
\begin{align*}
g(A) \;:=\; \,&-D \, A_{11} \, \log A_{11} - D \, A_{21} \, \log A_{21}
\notag \\
&-U \, A_{12} \, \log A_{12} - U \, A_{22} \, \log A_{22}
\,,
\end{align*}
is Lipschitz continuous on closed subdomains. In particular, for all $
A,A' \in B_\varepsilon(T_*) $ holds
\begin{equation}
\label{g is Lipschitz}
\abs{g(A) - g(A')} \;\leq\; k_2 \, \norm{A - A'}\,,
\end{equation}
provided that $ \varepsilon > 0 $ is small enough that the closure of
the ball $ B_\varepsilon(T_*) = \sett{A \in \R^{2\times
    2}\,:\,\norm{A-T_\ast} < \varepsilon} $ is contained in $ (0,1)^{2
  \times 2} $, and the finite constant $ k_2 = k_2(u,v,\varepsilon) $
is large enough.

Third, by a direct calculation we see that $ g $ satisfies
\begin{equation*}
\mathscr{H} \pb{X^{(j)}} \;=\; g(T^j)
\quad\text{and}\quad 
\tH(X_1) \;=\; g(T_\ast)\,.
\end{equation*}

Therefore, by expressing the difference of the recursion
relations~\eqref{the recursion relation of upper bounds} and~\eqref{the
  recursion relation of lower bounds} with these identities and using
the trivial bound $ p_j \leq 1 $, we get
\begin{align}
\mathscr{U}_j - \mathscr{L}_j \;&=\; \frac{p_{j-1}}{D} \, \qb{ \mathscr{H} \pb{X^{(j)}} - \tH (X_1)} \notag \\
&\leq\; \frac{1}{D} \, \qb{ g(T^j) - g(T_\ast)}\,.
\label{first convergence estimate}
\end{align}
If $ j $ is large enough the estimates~\eqref{g is Lipschitz}
and~\eqref{matrix convergence} can be combined to yield
\begin{equation*}
g\pb{T^j} - g(T_\ast) 
\;\leq\; 
k_2 \, \normb{T^j - T_*} 
\;\leq\; 
k_1 \, k_2 \, \abs{1 - u - d}^j \,,
\end{equation*}
which together with~\eqref{first convergence estimate} completes the
proof.
\end{proof}

Finally some remarks about convergence. From the theorem it is clear
that if $u+d \approx 1$ the convergence is fast. Indeed, if $u+d = 1$
the first order terms $\mathscr{L}_1 = \mathscr{U}_1$ are exact. This
can also be seen directly: We have $u = U$, $d = D$, so that $T = T^2
= T_*$ and therefore $\mathscr{H}(X) = \mathscr{H}\pb{X^{(j)}} =\tH
(X_1)$. On the other hand, the convergence becomes slower if $ u,d
\approx 0 $ or $ u,d \approx 1 $. The limiting case $u = d = 0$
corresponds to a static network and $u = d = 1$ is physically
meaningless, which is also why we excluded both cases from our
discussion.


\section{Conclusion}

In a dynamic network information about connectivity must be sent through
the network regularly. This connection state overhead consumes the
available bandwidth of the network. It is therefore natural to ask what is the smallest possible (in the context of information theory) bandwidth required for the connection state overhead.
In this work we provide the answer in the special
case of a simple linear network model:
As a main result we have presented an exact and rapidly converging
series expression for the best achievable overhead data rate.

We have only considered a linear network model. However, the results
derived here are also applicable to the case of a tree with the
connectivity information at each node being whether or not it is
connected to the root, since this model is fully equivalent to the
one-dimensional network.

The generalisation of our results to linear networks with more general
links that have a larger state space is probably possible by using the
same or very similar techniques as here. However, the most interesting
generalisations, such as more complex network topologies, seem to pose a far greater challenge.

\appendices

\section{An effective algorithm for computing $r_j$} 
\label{sect: calculation of p_j} 
A ``brute force'' computation of $ p_j $ is too complex to be of any
practical use if $j > 2$. 
We present here a more convenient method. 
The result is a simple recursive algorithm for calculating 
$ r_j $.    
The probabilities $ p_j $ can then be computed from
\begin{equation*}
p_j = 1 - r_1 - \dots - r_j.
\end{equation*}

For $j \in \N$ we have
\begin{align}
r_j \;=\; &p_{j-1} - p_j \notag \\ 
\;\sim\; 
&\tP \qB{M_t > \max\{M_{t-1}, \dots, M_{t - j + 1}\}, M_t \leq M_{t - j}}
\notag \\ 
\label{computation of r_j}
\;=\; 
&\begin{aligned}[t]
\sum_{m = 0}^\infty \;\tP \qb{M_t = m}\, \tP \Bigl[\, &M_{t - j} \geq m,M_{t - j + 1} < m, \cdots \\ 
&\cdots, M_{t - 1} < m\, \Big|\, M_t = m\, \Bigr]
\,. 
\end{aligned}
\end{align}
Define the new random variable
\begin{equation*}
Z^{(m)}_t \;:=\; \prod_{x = 1}^m \; Z_t(x)\,,
\end{equation*}
so that
\begin{equation*}
\hb{Z^{(m)}_t = 0} \;=\; \{M_t < m\}\,. 
\end{equation*}
Then we get from above
\begin{align}
r_j &\;\sim\; \sum_{m = 0}^\infty \; 
\begin{aligned}[t] 
\tP \Big[&Z^{(m)}_{t - j} = 1, Z^{(m)}_{t - j + 1} = \dots = Z^{(m)}_{t - 1} = 0 
\Big| \\    
&Z^{(m)}_t = 1, Z_t(m + 1) = 0 \Bigr] \, \tP \qb{M_t = m}\end{aligned}
\notag \\ \label{r as a sum over m}
&\;\sim\; D \sum_{m = 0}^\infty  \; r_j^{(m)}\, U^m 
\,,
\end{align}
where we have used~\eqref{prob distribution of M} and $ r_j^{(m)} $ is
the limit
\begin{equation*}
\lim_{t \to \infty} 
\tP \qcB{Z^{(m)}_{t - j} = 1, Z^{(m)}_{t - j + 1} = \dots = Z^{(m)}_{t - 1} = 0}
{Z^{(m)}_t = 1}
\,.
\end{equation*}
The above discussion is meaningless if $m = 0$; from
\eqref{computation of r_j}, however, we see that we must define
\begin{equation*}
r_j^{(0)} \;:=\;
\begin{cases}
1\,, &j = 1 \,,
\\
0\,, &j > 1 \,,
\end{cases}
\end{equation*}
for \eqref{r as a sum over m} to hold.

Define now for $j \in \N$
\begin{equation*}
q_j^{(m)} \;:=\; \lim_{t \to \infty} 
\tP \qcb{\,Z^{(m)}_{t - j} = 1}{Z^{(m)}_t = 1\,}
\,.
\end{equation*}
For the following we note that the process obtained from $X$ by
reversing the time is also a Markov process with transition
probabilities identical\footnote{We use here the fact that the links
  are distributed according to the stationary distribution at all
  times.} to those of $X$; for example $\tP \qc{X_{t - 1} = 1}{X_t =
  0} \,=\, u$. Thus
\begin{equation*}
q_j^{(m)} \;=\; \overline{d}_j^m\,.
\end{equation*}

The recursion relation for $r_j^{(m)}$ arises as follows. We rewrite
$q_j^{(m)}$ by decomposing the event $\hb{Z^{(m)}_{t - j} = 1,
  Z^{(m)}_t = 1}$: By successively conditioning on the values of
$Z^{(m)}_{t - i}$, $i = 1, \dots, j$, we get
\begin{align*}
\bigl\{ Z^{(m)}_{t - j} = 1, Z^{(m)}_t = 1\bigr\} \;= \; \sum_{i = 1}^j \; 
\Bigl\{Z^{(m)}_{t - j} = 1, Z^{(m)}_{t - i} = 1&, \\ 
Z^{(m)}_{t - i + 1} = \dots = Z^{(m)}_{t - 1} = 0, Z^{(m)}_t = 1&\,\Bigr\}\,,
\end{align*}
where the sum means a union of disjoint events. Taking the probability
measure of both sides and using Markovity\footnote{Note that if
  $Z^{(m)}_t = 1$ then \emph{all} of the relevant first $m$ links of
  $\ul{Z}_{t}$ are known (to equal 1).} of the time-reversed $X$
process we have
\begin{equation*}
q^{(m)}_j \;=\; \sum_{i = 1}^j \; q_{j - i}^{(m)} \, r_i^{(m)}\,,
\end{equation*}
which gives
\begin{equation} 
\label{recursion relation for r_j^m}
r_j^{(m)} \;=\; \overline{d}_j^m 
- \sum_{i = 1}^{j - 1} \overline{d}_{j - i}^m \, r^{(m)}_i
\,.
\end{equation}
This is the desired recursion relation expressing $r_j^{(m)}$ as a
function of $r_1^{(m)}, \dots, r_{j-1}^{(m)}$. Using $r^{(m)}_1 =
\overline{d}^m$ we may therefore find $r^{(m)}_j$.

We summarise the results:
\begin{lemma}
The quantity $r_j$, $j \in \N$, may be computed from
\begin{equation*}
r_j \;=\; D \, \sum_{m = 0}^\infty \; r_j^{(m)} \, U^m \,,
\end{equation*}
where $r^{(m)}_j$, $m \in \N_0$, satisfies the recursion relation
\begin{subequations}
\begin{align*}
r_j^{(m)} &\;=\; \overline{d}_j^m - \sum_{i = 1}^{j - 1} \overline{d}_{j - i}^m \, r^{(m)}_i\,,
\\
r^{(m)}_1 &\;=\; \overline{d}^m \,.
\end{align*}
\end{subequations}
\end{lemma}

As an example, we compute $r_1$, $r_2$ and $r_3$:
\begin{align*}
r_1 \;=\;
&D \, \sum_{m = 0}^\infty \; \overline{d}^m \, U^m \;=\; D \, \frac{1}{1 - \overline{d} \, U} \,,
\\
r_2 \;=\;
&D \, \sum_{m = 0}^\infty \; \pB{\overline{d}_2^m - \overline{d}^{2m}} \, U^m \\
\;=\; 
&D \, \qbb{\frac{1}{1 - \overline{d}_2 \, U} - \frac{1}{1 - \overline{d}^2 \, U}} \,,
\\
r_3 \;=\; 
&D \, \sum_{m = 0}^\infty \; \pB{\overline{d}_3^m -  2 \, \overline{d}^m \, \overline{d}_2^m + \overline{d}^{3m}} U^m \\
\;=\; 
&D \, \qbb{\frac{1}{1 - \overline{d}_3 \, U} - \frac{2}{1 - \overline{d} \, 
\overline{d}_2 \, U} + \frac{1}{1 - \overline{d}^3 \ U}}
\,.
\end{align*}


\section{Proof of Lemma \ref{lemma: proof of steps}} 
\label{app: proof of steps}
The proof involves deriving equalities for conditional probabilities.
These then induce equalities of the conditional entropies according to
the following lemma.
\begin{lemma} \label{lemma: prob -> entropy}
Let $X,Y$ be random variables, $\phi$ a function on the range of $Y$, and suppose that, for all $x,y$,
\begin{equation*}
\tP [X = x \,|\, Y = y] \;=\; \tP\bigl[X = x \,|\, Y \in \phi^{-1} \pb{\phi(y)}\bigr]\,. 
\end{equation*}
Then
\begin{equation*}
\tH(X \,|\, Y) \;=\; \tH(X \,|\, \phi(Y))
\,.
\end{equation*}
\end{lemma}
\begin{proof}
  The proof is based on writing out the definition of the conditional
  entropy $\tH (X \,|\, Y)$, rewriting the sum $\sum_{x, y} \,
  (\genarg)$ as $\sum_{x,s} \sum_{y \,:\, \phi(y) = s} \, (\genarg)$
  and using the assumption. We omit further details.
\end{proof}

\begin{proof}[Lemma \ref{lemma: proof of steps}]
Let us begin with (i). The conditioning event is 
\begin{equation*}
A^\cmpl \;=\; \bigl\{ M_t \leq \abs{\ul{M}_{t-1}^{j}} \bigr\}
\,.
\end{equation*}
Let $\ul{m} \in \N^j $ and define 
\[
i(\ul{m}) \;:=\; \min\hb{k \in \{1,  \dots, j\} \,:\, m_{t - k} = \abs{\ul{m}}}\,.
\] 
Let furthermore
$\ul{z}', \ul{z}'' \in \{0,1\}^{\N}$ be chosen so that
$\varphi(\ul{z}') = m_{t - j}$ and $\varphi(\ul{z}'') = m_{t -
  i(\ul{m})}$, where $\varphi$ is a deterministic function that gives
$ M_t $ as a function of $ \ul{Z}_t $. Then we have, for $m \in \N$
and $\ul{z} \in \{0,1\}^{\N}$,
\begin{align*}
\tP \Bigl[\,&M_t = m \,\Big|\, \ul{M}_{t - 1}^{(j)} = \ul{m}, 
\ul{Z}_{t -j - 1} = \ul{z}, M_t \leq \abs{\overline{m}}\,\Bigr] 
\\
\overset{\text{(a)}}{=}\; 
\tP \Bigl[\,&Z_t(1) = \dots= Z_t(m) = 1, Z_t(m + 1) = 0 \;\Big| \\
&\ul{M}_{t - 1}^{(j)} = \ul{m}, \ul{Z}_{t -j - 1} = \ul{z}, M_t \leq \abs{\overline{m}}, \\
&Z_{t - i(\ul{m})}(1) = \dots = Z_{t - i(\ul{m})}(1) = 1,\\ &Z_{t - i(\ul{m})}(\abs{\ul{m}} + 1) = 0\;\Bigr]
\\
\overset{\text{(b)}}{=}\; 
\tP \Bigl[\,&Z_t(1) = \dots= Z_t(m) = 1, Z_t(m + 1) = 0 \,\Big|
\notag \\
&\ul{M}_{t - 1}^{(j)} = \ul{m}, M_t \leq \abs{\ul{m}}, Z_{t - i(\ul{m})}(1) = \cdots \\
&\dots = Z_{t - i(\ul{m})}(1) = 1, Z_{t - i(\ul{m})}(\abs{\ul{m}} + 1) = 0\,\Bigr]
\notag \\
=\;\tP \Bigl[\,&M_t = m\,\Big|\,
\ul{M}_{t - 1}^{(j)} = \ul{m}, M_t \leq \abs{\ul{m}}\,\Bigr]
\notag \\
\overset{\text{(c)}}{=} \; 
\tP \Bigl[\,&Z_t(1) = \dots= Z_t(m) = 1, Z_t(m + 1) = 0 \,\Big|\\
&\ul{M}_{t - 1}^{(j)} = \ul{m}, M_t \leq \abs{\ul{m}}, \ul{Z}_{t - i(\ul{m})} = \ul{z}''\,\Bigr]
\\
\overset{\text{(d)}}{=} \; \tP \Bigl[\,&Z_t(1) = \dots= Z_t(m) = 1, Z_t(m + 1) = 0 \,\Big| \\
&\ul{M}_{t - 1}^{(j)} = \ul{m}, M_t \leq \abs{\ul{m}}, \ul{Z}_{t - i(\ul{m})} = \ul{z}'', \ul{Z}_{t - j} = \ul{z}' \,\Bigr]
\\
\overset{\text{(e)}}{=} \; \tP \Bigl[\,&M_t = m \,\Big|\, \ul{M}_{t - 1}^{(j)} = \ul{m}, M_t \leq \abs{\ul{m}}, \ul{Z}_{t - j} = \ul{z}'\,\Bigr]
\notag \\
=\; \tP \Bigl[\,&M_t = m \,\Big|\, \ul{M}_{t - 1}^{(j - 1)} = \ul{m}^{(j - 1)}, M_t \leq \abs{\ul{m}}, \ul{Z}_{t - j} = \ul{z}'\,\Bigr]
\,,
\end{align*}
where $\ul{m}^{(j - 1)}$ denotes the $j - 1$ first components of
$\ul{m}$; (a) follows from rewriting the conditions $M_t = m$, $M_{t -
  i(\ul{m})} = \abs{\ul{m}}$; (b) from Markovity, independence and the
fact that $m \leq \abs{\ul{m}}$; (c) from independence and $m \leq
\abs{\ul{m}}$; (d) from Markovity; and (e) from independence and $m
\leq \abs{\ul{m}}$. Then the assertion follows from Lemma \ref{lemma:
  prob -> entropy} by choosing the functions $\phi_1\pb{\ul{m}^{(j -
    1)}, \ul{z}', \ul{z}} := \pb{\ul{m}^{(j - 1)},
  \varphi(\ul{z}'),\ul{z}}$, $\phi_2\pb{\ul{m}^{(j - 1)}, \ul{z}',
  \ul{z}} := \pb{\ul{m}^{(j - 1)}, \varphi(\ul{z}')}$, and
$\phi_3\pb{\ul{m}^{(j - 1)}, \ul{z}', \ul{z}} := \pb{\ul{m}^{(j - 1)},
  \ul{z}'}$.

To prove (ii) choose $\ul{m}$, $\ul{z}$ and $\ul{z}'$ as above and write
\begin{align*}
\tP \bigl[\, &M_t > \abs{\ul{m}} \,\big|\, \ul{M}_{t-1}^{(j)} = \ul{m}, 
\ul{Z}_{t-j-1} = \ul{z} \,\bigr] 
\\ \overset{\text{(a)}}{=}\; 
\tP \bigl[\, &Z_t(1) = \dots = Z_t(\abs{\ul{m}} + 1) = 1 \,\big|\,\\ 
&\ul{M}_{t-1}^{(j)} = \ul{m}, \ul{Z}_{t-j-1} = \ul{z}\,\bigr]
\\ \overset{\text{(b)}}{=}\; 
\tP \bigl[\,&Z_t(1) = \dots = Z_t(\abs{\ul{m}} + 1) = 1 \,\big|\, 
\ul{M}_{t-1}^{(j)} = \ul{m} \,\bigr]
\\ \overset{\text{(c)}}{=}\; 
\tP \bigl[\,&Z_t(1) = \dots = Z_t(\abs{\ul{m}} + 1) = 1 \,\big|\,\\ 
&\ul{M}_{t-1}^{(j)} = \ul{m}, \ul{Z}_{t-j} = \ul{z}' \,\bigr]
\\ =\; \tP \bigl[\,&M_t > \abs{\ul{m}} \,\big|\, 
\ul{M}_{t-1}^{(j-1)} = \ul{m}^{(j-1)}, \ul{Z}_{t-j} = \ul{z}' \,\bigr]
\,,
\end{align*}
where (a) follows from rewriting $M_t > \abs{\ul{m}}$; (b) and (c)
from independence and Markovity (the full details are exactly as above
using the index variable $i(\ul{m})$).

The proofs of (iii), (iv) and (v) are almost identical; we only show
(iv). Let $m$, $\ul{m}$ and $\ul{z}$ be as above. First note that
under the conditions $\ul{M}_{t - 1}^{(j)} = \ul{m}$ and $M_t >
\abs{\ul{m}}$ there is a bijective map between $M_t$ and
$\wti{M}_t(\abs{\ul{m}} + 2)$:
\begin{equation*}
M_t \;=\; \wti{M}_t(\abs{\ul{m}} + 2) \,+\, \abs{\ul{m}} \,+\, 1\,,
\end{equation*}
so that
\begin{align}
&\tH \pcb{M_t}{\ul{Z}_{t - j - 1} \,:\, \hb{\ul{M}_{t - 1}^{(j)} = \ul{m}, M_t > \abs{\ul{m}}}}
\label{conditioned entropy translated}
\\ 
&=\;
\tH \pcb{\wti{M}_t(\abs{\ul{m}} + 2)}{\ul{Z}_{t - j - 1} \,:\, \hb{\ul{M}_{t - 1}^{(j)} = \ul{m}, M_t > \abs{\ul{m}}}} \,.
\notag
\end{align}
Now 
\begin{align*}
\tP \Bigl[\,&\wti{M}_t(\abs{\ul{m}} + 2) = m \,\Big|\, 
\ul{M}_{t - 1}^{(j)} = \ul{m}, \ul{Z}_{t - j - 1} = \ul{z}, 
M_t > \abs{\ul{m}}\Bigr]
\\ \overset{\text{(a)}}{=}\; 
\tP \Bigl[\,&\wti{M}_t(\abs{\ul{m}} + 2) = m \,\Big|\,
\ul{M}_{t - 1}^{(j)} = \ul{m}, \ul{Z}_{t - j - 1} = \ul{z},Z_t(1) = \\
&\cdots= Z_t(\abs{\ul{m}} + 1) = 1\,\Bigr]
\\
\overset{\text{(b)}}{=}\; \tP \Bigl[\,&\wti{M}_t(\abs{\ul{m}} + 2) = m 
\,\Big|\, \ul{Z}_{t - j - 1} = \ul{z}\,\Bigr] \,,
\end{align*}
where (a) follows from rewriting the condition $M_t > \abs{\ul{m}}$,
and (b) from independence.  Now by Lemma \ref{lemma: prob -> entropy}
and \eqref{conditioned entropy translated} we get
\begin{align*}
&\tH \pcb{M_t}{\ul{Z}_{t - j - 1} \,:\, \{ \ul{M}_{t - 1}^{(j)} = \ul{m}, 
M_t > \abs{\ul{m}}\} } 
\\
&=\; \tH \pcb{\wti{M}_t(\abs{\ul{m}} + 2)}{\ul{Z}_{t-j-1}}
\;\sim\; \tH \pc{M_t}{\ul{Z}_{t - j - 1}}\,,
\end{align*}
where the last step follows from translation invariance. Therefore
\begin{align*}
&\tH \pcb{M_t}{\ul{M}_{t - 1}^{(j)}, \ul{Z}_{t-j-1} \,:\, A}
\\
&= \; \sum_{\ul{m}}\! 
\begin{aligned}[t]
&\tP \qcb{\ul{M}_{t - 1}^{(j)} = \ul{m}}{A} \\ 
\cdot\;&\tH \pcb{M_t}{\ul{Z}_{t - j - 1} \,:\, \hb{\ul{M}_{t - 1}^{(j)} = \ul{m}, M_t > \abs{\ul{m}}}}
\end{aligned}
\\
&\sim\; \tH \pcb{M_t}{\ul{Z}_{t-j-1}} \sum_{\ul{m}} \; \tP \qcb{\ul{M}_{t - 1}^{(j)} = \ul{m}}{A}
\\
&= \; \tH \pcb{M_t}{\ul{Z}_{t-j-1}}\,.
\end{align*}
\end{proof}


\section*{Acknowledgements}

The authors are grateful for Ari Hottinen, Antti Kupiainen, Paolo
Muratore-Ginanneschi and Olav Tirkkonen for fruitful discussions
during the preparation of this paper.


\end{document}